\documentclass{iaus}
\usepackage{graphicx}

\title[Asteroseismic diagram] 
{Asteroseismic study of solar-like stars: A method of estimating
stellar age}

\author[Y. K. Tang, S. L. Bi \& N. Gai]   
{Y. K. Tang$^{1,3}$%
  \thanks{Present address:},
 S. L. Bi$^{2,1}$ \and N. Gai$^{1,3}$}

\affiliation{$^1$National Astronomical Observatories/Yunnan
Observatory, Chinese Academy of Sciences, Kunming 650011,
   P. R. China \break email: bisl@bnu.edu.cn; tangyanke@ynao.ac.cn\\[\affilskip]
$^2$Department of Astronomy Beijing Normal University, Beijing
100875,
  P. R. China \\[\affilskip]
$^3$Graduate School of the Chinese Academy of Sciences, Beijing
100039, P. R. China}

\pubyear{2008}
\volume{252}  
\pagerange{ -- }
\date{?? and in revised form ??}
 \jname{The art of modelling stars in the 21st
century} \editors{Licai Deng, K.L. Chan \& C. Chiosi, eds.}
\begin{document}

\maketitle

\begin{abstract}
Asteroseismology, as a tool to use the indirect information
contained in stellar oscillations to probe the stellar interiors, is
an active field of research presently. Stellar age, as a fundamental
property of star apart from its mass, is most difficult to estimate.
In addition, the estimating
of stellar age can provide the chance to study the time evolution of
astronomical phenomena. In our poster, we summarize our previous
work and further present a method to determine age of low-mass
main-sequence star.

\end{abstract}

\firstsection 
\section{Introduction}
Due to the frequencies of these oscillations depend on density,
temperature, gas motion, and other properties of the stellar
interior, it can take the window to ``see" the interior of stars and
help us to know the stellar internal structure and understand the
stellar evolution. With the advance of observational technique, 
several stars have been detected the solar-like oscillations. Using
the latest asteroseismic data, we reconstruct the model of $\alpha$
Cen B and 70  Ophiuchi A (Tang et al. 2008a, 2008b). In additional,
\cite{Bi} have performed preliminary seismological analysis of two
MOST targets. 
\section{Using asteroseismic diagram to estimate stellar age}\label{sec:greenfun}
Following the asymptotic formula for the frequency $\nu_{n,l}$ of a
stellar $p$-mode of order $n$ and degree $l$ was given by
Tassoul(1980):
\begin {equation}
\nu_{n,l}\simeq(n+\frac{l}{2}+\epsilon)\nu_{0}-[Al(l+1)-B]\nu_{0}^{2}\nu_{n,l}^{-1},
\end {equation}
where, $\nu_{0}$ and $A$ are related to the run of sound speed.
Based on some quantities as diagnostic purpose to probe the stellar
interior proposed by some authors (Christensen-Dalsgaard 1988; Gough
2003; et al.), like $ \delta\nu_{n,l} = \nu_{n,l}-\nu_{n-1,l+2}$, we
definite another quantity (Tang et al. 2008)

\begin {equation}
r_{01}=\frac{\langle\delta\nu_{0,2}\rangle}{\langle\delta\nu_{1,3}\rangle}.
\end {equation}

The $r_{01}$ comes from the perturbation to the gravitational
potential, neglected in equation (2.1), which affects modes of the
lowest degrees most strongly and which probably increases with
evolution due to the increasing central density for modes of the
lowest degrees which penetrate most deeply and hence affect
$\delta\nu_{0,2}$ more than $\delta\nu_{1,3}$, leading to the
dependence of $r_{01}$ on age. We compute some models with initial
heavy metal abundance $Z_{i}=0.02$, initial helium abundance
$Y_{i}=0.28$ and mixing-length parameter $\alpha=1.7$ using the Yale
stellar evolution code (YREC; Guenther et al. 1992) and analyze the
pulsation of low degree $p$-modes ($l= 0 - 3$) for selected models
in each given mass is implemented using the Guenther's pulsation
code under the adiabatic approximation (Guenther 1994).Considering
the $r_{01}$ is tightly correlated with age,
 we construct another asteroseismic diagram shown in Fig. 1b, based on the values
obtained from the above computation. Interestingly, the age of star
can be marked in this asteroseismic diagram Fig. 1b. It is
convenient to obtain the stellar important parameters: the mass and
the age.

\section{Discussion}
1. The ($\langle\Delta\nu\rangle$, $r_{01}$) diagram as a new
asteroseismic diagnostic tool can estimate the mass and the age of
solar-like stars. The virtue is that the age of stars can be marked
in the diagram, so we can obtain the mass and age directly.

2. We will discuss the effects of the assumed initial abundance of
helium and the mixing length parameter on the asteroseismic diagram
in future work.
\begin{figure}
\includegraphics[height=1.5in,width=2.5in,angle=0]{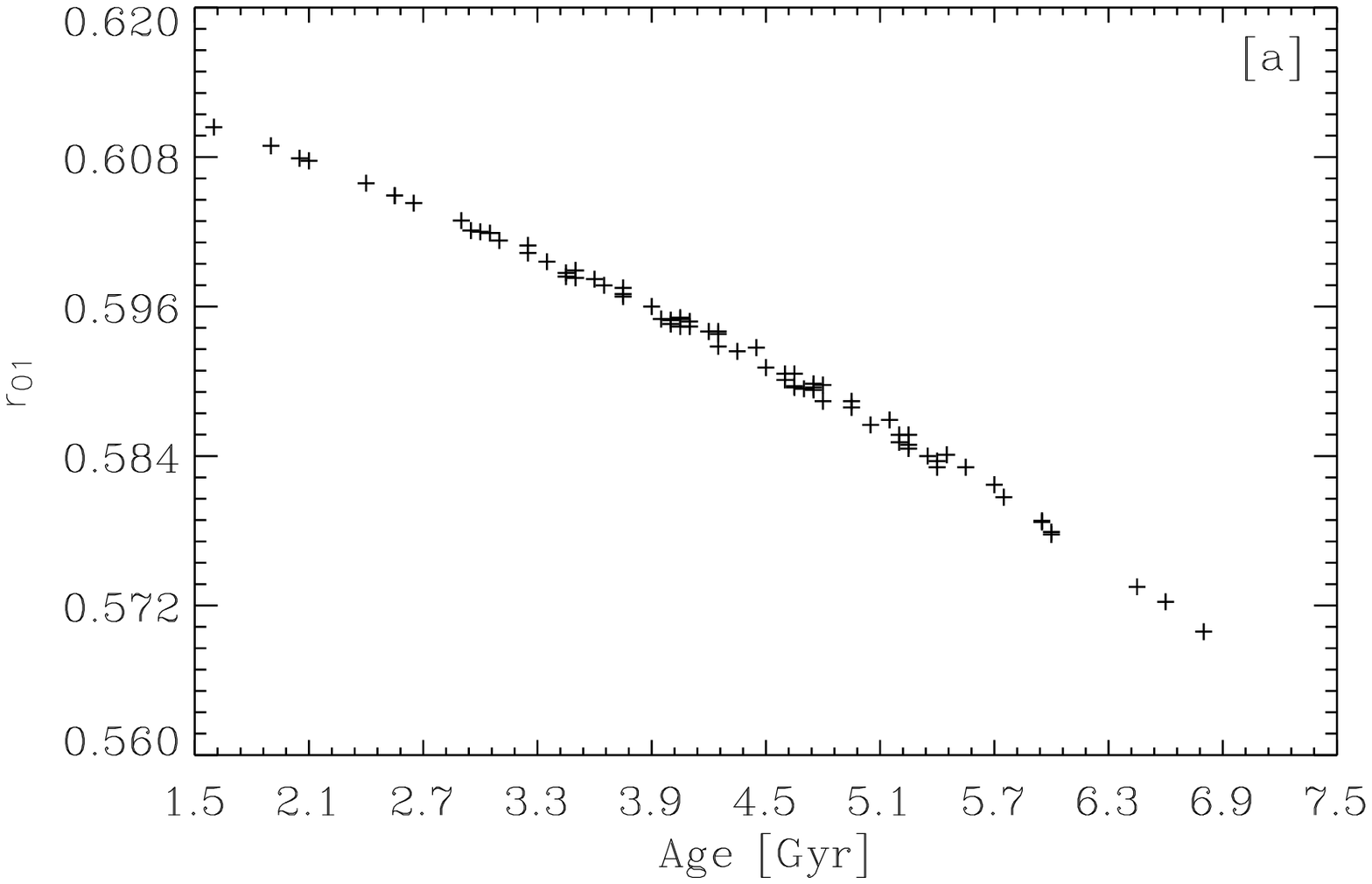}%
\includegraphics[height=1.5in,width=2.5in,angle=0]{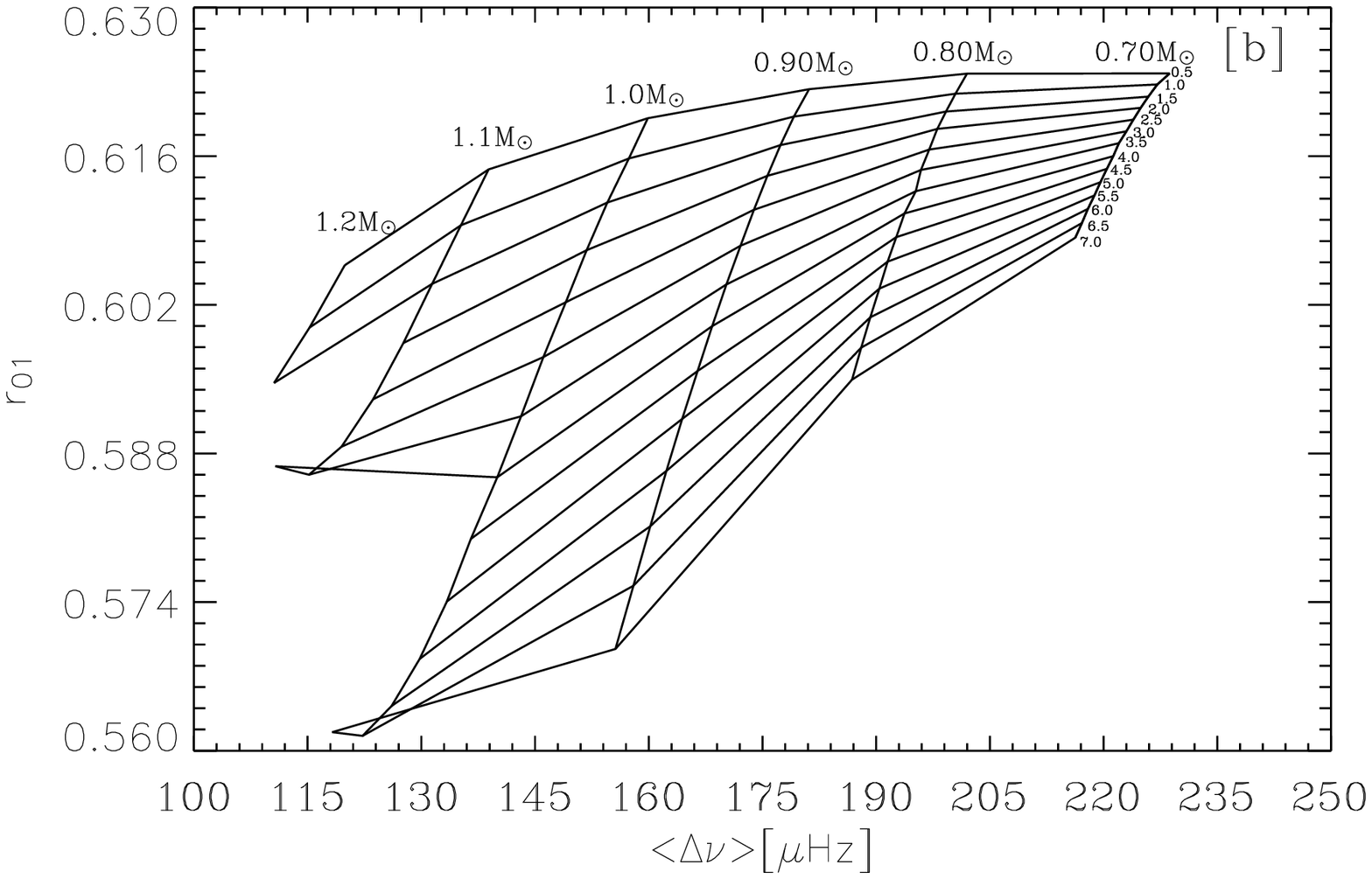}
  \caption{[a]: The ratio of small separations adjacent in $l$ vs. age for
each of 129 stellar models described in \cite{Tanga}. [b]:
($\langle\Delta\nu\rangle$, $r_{01}$) diagram for stellar models.
The vertical lines are evolutionary tracks, labeled by the mass in
the top, whereas the transverse lines are isopleths with constant
age, labeled by the age from 0.5 Gyr to 7.0 Gyr increasing with 0.5
(unit is Gyr )}\label{fig:contour}
\end{figure}

\begin{acknowledgments}
This work was supported by The Ministry of Science and Technology of
the People's republic of China through grant 2007CB815406, and by
NSFC grants 10173021, 10433030, 10773003, and 10778601.
\end{acknowledgments}

%
%

\end{document}